\documentclass[preprintnumbers,amsmath,amssymb,floatfix,11pt,prd,onecolumn,
superscriptaddress,nofootinbib]{revtex4}
\usepackage{latexsym}
\usepackage{epsfig}
\usepackage{epstopdf}
\usepackage{amssymb,bigints}

\begin{document}

\title{\bf Strong Gravitational Lensing by Kiselev Black Hole}

\author{Azka Younas}
\affiliation{Department of Physics, School of Natural Sciences (SNS), National University
	of Sciences and Technology (NUST), H-12, Islamabad, Pakistan}

\author{Mubasher Jamil}
\email{mjamil@sns.nust.edu.pk}\affiliation{Department of Mathematics, School of Natural Sciences (SNS), National University of Sciences and Technology (NUST), H-12, Islamabad, Pakistan}
\author{Sebastian Bahamonde}
\email{sebastian.beltran.14@ucl.ac.uk}\affiliation{Department of Mathematics,University College London,
	Gower Street, London, WC1E 6BT, United Kingdom}
\author{Saqib Hussain}
\affiliation{Department of Physics, School of Natural Sciences (SNS), National University
	of Sciences and Technology (NUST), H-12, Islamabad, Pakistan}

\begin{abstract}
{\centering \bf Abstract:} We investigate the gravitational lensing
scenario due to Schwarzschild-like black hole surrounded by
quintessence (Kiselev black hole). We work for the special case of Kiselev black hole where we take the state parameter $w_{q}=-\frac{2}{3}$. For the detailed derivation and analysis of the bending
angle involved in the deflection of light, we discuss three special cases of Kiselev black hole: nonextreme, extreme and naked
singularity. We also calculate the approximate bending angle and compare it
with the exact bending angle. We found the relation of bending angles in the decreasing order as: naked singularity, extreme Kiselev black hole, nonextreme Kiselev black hole and Schwarzschild black hole. In the weak field
approximation, we compute the position and total magnification of
relativistic images as well.
\end{abstract}

 \maketitle

\section{Introduction}
Gravitational lensing (GL) signifies the deflection of electromagnetic waves. Light propagates in empty space along a straight line. The well-known theory of General Relativity (GR) predicts that light will be bent if an object with a certain gravitational field is interposed in the light path. In literature, GL has been used to study highly redshifted galaxies, quasars, supermassive black holes, exoplanets, dark matter candidates, primordial gravitational wave signatures, etc., \cite{hist}.
In $1801$, Soldner was the first person who calculated the bending angle of light by using Newtonian Mechanics \cite{sold}. In $1911$, Einstein derived the same Soldner's result by using the equivalence principle and Minkowski metric, unaffected by gravity \cite{Eins}. This marks the beginning of our modern understanding of GL. In 1915, Einstein derived the new solar light deflection angle that was double from the previous value due to the effect of the spacetime curvature \cite{Dyson}. Eddington in $1919$, confirmed the prediction of Einstein during the solar eclipse
\cite{Eddin}.
 In $1937$, Zwicky estimated the gravitational lens effect can be observed \cite{Zwick}.
 In $1979$, Walsh, Weymann and Carswell used Zwicky's work and discovered the first example of GL in which they obtained the first multiple images of a binary quasar (QSO $0957+561$)
 \cite{Walsh}.

In 1959, Darwin calculated the light deflection angle due to a strong gravitational field using the Schwarzschild metric \cite{Darwin}. Another significant work involved the deflection angle and intensities for the images formed due to the Schwarzschild black hole in terms of elliptic integrals of the first kind \cite{Ohani}. Considering the Schwarzschild black hole for the strong GL, Virbhadra and Ellis obtained the lens equation and introduced a method to calculate the bending angle. They also studied the lensing problem for the galactic supermassive black hole numerically \cite{Virbh}. While studying GL with the Schwarzschild black hole in the strong field limit, the bending angle was also evaluated analogous to the weak field limit. Besides the weak field limit of relativistic images, magnifications and critical curves formulas were also formulated \cite{Bozza}. Bozza treated the strong lensing phenomenon by a spherically symmetric black hole, where an infinite sequence of higher order images are formed \cite{bozi} and later on extended for a spinning black hole \cite{boz1}. One of the first important studies about a cosmological constant relativistic bending angle was done by Rindler and Ishak where they showed that for a Schwartzschild de Sitter geometry, the cosmological constant does not contribute to the bending angle \cite{rindler}. Another important application of relativistic bending angle techniques were used to determine a limit in the cosmological constant by using the bending of light through galaxies and clusters of galaxies \cite{ishak}.

About two decades ago, a very important astronomical observation (using Supernovae type Ia) suggested that the Universe is in a state of an accelerated expansion \cite{Riess,perl}. This study was a revolution in physics and the dark energy was named to be responsible for this accelerating scenario. Cosmologists proposed different models in order to explain this strange behaviour of the Universe such as the $\Lambda$CMD model (with a state parameter of $w=-1$) or dynamic scalar fields \cite{cosmoC,DYs}. The former uses the old idea of a cosmological constant introduced by Einstein several years ago but in a completely different way,\footnote{Einstein introduced a cosmological constant in his field equations to obtain a static universe. After some observations that suggested that the Universe is expanding, Einstein thought that this constant was the worst mistake in his life. However, nowadays, this constant has been taken into account but using another physical interpretation related with dark energy} now interpreted like a reason to support the dark energy. However, this model has some problems like the so-called ``cosmological constant problem" where the value of the cosmological constant differs about $10^{120}$ orders of magnitude from the empirical value \cite{Cosmproblem}. The second candidate for dark energy is a dynamic scalar field such as quintessence, phantoms, k-essence, etc. \cite{Carroll,Caldwell,Armen}. Generally, a quintessence model has a state parameter $w(t)=p(t)/\rho(t)$, where $p(t)$ is the pressure and $\rho(t)$ is the energy density that varies with time depending on the energy potential $V(\Phi)$ and scalar field $\Phi$. In addition, it is important to mention that the quintessence field is minimally coupled to gravity and the potential energy decreases as the field increases. This model is the simplest case without having theoretical problems like Laplacian instabilities or ghosts. For a more detailed review of the quintessence, see \cite{Rashmi,Copeland,Sfer}

One important solution related to the quintessence model was discovered by Kiselev \cite{Kisel}. The former solution physically describes a spherically symmetric and static exterior spacetime filled with a quintessence field, hence a nonvacuum solution. The Kiselev obtained the Schwarzschild-like and Reissner-Nordstr\"{o}m-de Sitter BH's solutions surrounded by the quintessence at the range of state parameter $-1<w_{q}<-\frac{1}{3}$,  the Universe will accelerate with the quintessence, where $w_{q}$ is the ratio of pressure and energy density of quintessence. At $w_{q}=-1$, quintessence covers the cosmological constant $\Lambda$ term and corresponds to the case of dark energy, while $w_{q}<-\frac{1}{3}$, in a static coordinates quintessential state, reveals a de Sitter type outer horizon. In short, the solutions that corresponds to $-1<w_{q}<-\frac{1}{3}$ are asymptotically de Sitter. In this paper, we study the gravitational lensing due to a Kiselev black hole (KBH) where we choose the state parameter $w_{q}=-\frac{2}{3}$. Due to this value, the solution will be a Schwarzschild-like (netural) black hole surrounded by quintessence \cite{Kisel}. In this paper, we considered three possibilities for KBH: two distinct horizons (nonextreme), unique horizon (extreme black hole) and no horizon (naked singularity). From the astrophysical point of view,
it is a hard task to distinguish between the signatures and
properties of black hole and naked singularities, however, GL can provide distinguishing signatures \cite{joshi}.

The paper is structured as follows: In Sec. II, we study the
geodesics and effective potential for nonextreme and naked
singularity. In Sec. III, we discuss critical variables and equation
of path for photons and calculate the relations between closest
approach $r_{o}$ and impact parameter $b$. In Sec. IV, we derive the
bending angle in terms of elliptical integrals for both nonextreme
KBH and naked singularity for different values of quintessence
parameter $\sigma$ (discussed later) and then make a comparison with the bending angle for a
Schwarzschild black hole. In Sec. V, we study the geodesics and effective
potential for extreme KBH. In Sec. VI, we discuss critical variables
and the equation of path for photons and calculate the relationship
between the closest approach and impact parameter for the extreme
lensing scenario. In Sec. VII, we calculate the bending angle
in terms of elliptical integrals for an extreme Kiselev black hole (EKBH) at a fixed value of $\sigma$ and compare it with the Schwarzschild bending angle as
a reference. In Secs. VIII, IX, X, we use an alternative method for
finding the bending angle to study the relativistic images. Finally we discuss our results in Sec. XI. We adopt the
units $c = G = 1$.

\section{Basic Equations for Null Geodesics in Kiselev Spacetime}

The equation of state parameter $w_{q}$ for the quintessence scalar field $\Phi$ is given by
\begin{equation}\label{2aa}
w_{q}=\frac{p_{q}}{\rho_{q}}=\frac{\frac{1}{2}\dot{\Phi}^{2}-V(\Phi)}
{\frac{1}{2}\dot{\Phi}^{2}+V(\Phi)},
\end{equation}
where $p_{q}$ and $\rho_{q}$ are the pressure and energy density of the quintessence field defined in terms of the kinetic energy ($\frac{1}{2}\dot{\Phi}^{2}$) and potential energy $V(\Phi)$, respectively. Here, the overdot represents the differentiation with respect to cosmic time.


Based on the above point of view, the geometry of a static spherically symmetric black hole surrounded by the quintessence (or Kiselev spacetime) is given by \cite{Kisel}
\begin{eqnarray}\label{1}
ds^{2}&=& f(r)dt^{2}-\frac{1}{f(r)}dr^{2}-r^{2}d\theta^{2}-r^{2}\sin^{2}\theta
d\phi^{2},
\nonumber\\ \text{where}\hspace*{10mm}
\nonumber\\
f(r)&=& 1-\frac{2M}{r}-\frac{\sigma}{r^{3w_{q}+1}}.
\end{eqnarray}
Here $M$ is the mass of the black hole and $\sigma$ is the quintessence
parameter (normalization factor) that is related to the energy density as follows \cite{Kisel}:
\begin{equation}\label{1aa}
\rho_{q}=-\frac{\sigma}{2}\frac{3w_{q}}{r^{3(1+w_{q})}}.
\end{equation}
When $w_{q}$ approaches $-1$, the function $f(r)$ for the metric $(\ref{1})$ reduces to
\begin{equation}\label{3aa}
f(r)=1-\frac{2M}{r}-\sigma r^{2},
\end{equation}
which is the Schwarzschild-de-Sitter black hole spacetime. For this case, the lensing phenomenon has been studied by Bakala and others \cite{pbzt1,Kott,shu}.
In this paper, our focus is on the special case $w_{q}=-\frac{2}{3}$, which corresponds to the Schwarzschild-like black hole surrounded by quintessence. In this case the function $f(r)$ becomes
\begin{equation}\label{1a}
f(r)=1-\frac{2M}{r}-\sigma r, ~~~~~ \Big(0<\sigma<\frac{1}{8M}\Big),
\end{equation}
which can also be written as
\begin{equation}\label{2a}
f(r)=\frac{\sigma}{r}(r-r_{-})(r-r_{+}).\hspace{30mm}
\end{equation}
The metric $(\ref{1})$ becomes ill defined at
$r=0$, i.e., $(g_{00}\rightarrow\infty)$  which gives a curvature
singularity. For $f(r)=0$, we get two fixed values of $r$, namely
\begin{equation}\label{2}
r_{+}=\frac{1+\sqrt{1-8M\sigma}}{2\sigma}, ~~~r_{-}=\frac{1-\sqrt{1-8M\sigma}}{2\sigma}. \hspace{20mm}
\end{equation}
The region $r=r_{-}$ corresponds to the black hole's event horizon while
$r=r_{+}$ represents the cosmological event horizon. Note that both
$r_{-}$ and $r_{+}$ are the two coordinate singularities in the
metric $(\ref{1})$. The coordinate singularities arise when
$0<\sigma<\frac{1}{8M}$. However when $\sigma>\frac{1}{8M}$, both
$r_{+}$ and $r_{-}$ become imaginary, giving a naked singularity.
When $\sigma=0$,  $r_{-}$ becomes the Schwarzschild BH's event horizon $r^{S}_{H}=2M$.

 The Lagrangian for a photon travelling in Kiselev
spacetime is given by
\begin{equation}\label{3}
  \mathcal{L}=\Big(1-\frac{2M}{r}-\sigma r\Big)\dot{t}^{2}-\frac{1}{1-\frac{2M}{r}-\sigma r}\dot{r}^{2}-r^{2}\dot{\theta}^{2}-r^{2}\sin^{2}\theta\dot{\phi}^{2}.
\end{equation}
Here dot represents the derivative with respect to $\lambda$ which is an affine parameter. We will work in an isotropic gravitational field, thus we can restrict the orbits of photons in the equatorial plane
$(\theta=\frac{\pi}{2})$. Hence, Eq. (\ref{3}) becomes
\begin{equation}\label{4}
  \mathcal{L}=\Big(1-\frac{2M}{r}-\sigma r\Big)\dot{t}^{2}-\frac{1}{1-\frac{2M}{r}-\sigma r}\dot{r}^{2}-r^{2}\dot{\phi}^{2}. \hspace{25mm}
\end{equation}
By using the Euler-Lagrange equations for null geodesics, we get
\begin{eqnarray}
 \dot{t}\equiv \frac{dt}{d\lambda}&=&\frac{E}{1-\frac{2M}{r}-\sigma r},\label{5}\\
\dot{\phi}\equiv\frac{d\phi}{d\lambda}&=&\frac{L}{r^{2}},\label{6}
\end{eqnarray}
where $E$ is the energy per unit mass and $L$ is the angular momentum per unit mass. Using the null condition of the 4-velocity
$g_{\mu\nu}u^{\mu}u^{\nu}=0$ (where $\mu,\nu=t,r,\theta,\phi$) and  $u^{\mu}=\frac{dx^{\mu}}{d\lambda}$ known as the 4-velocity we get the equation of motion for photons, that is
\begin{equation}\label{7}
  \dot{r}= L\sqrt{\frac{1}{b^{2}}-\frac{1}{r^{2}}\Big(1-\frac{2M}{r}-\sigma r\Big)},
   ~~~\textrm{where}~~~ b=\Big|\frac{{L}}{E}\Big|.\hspace{15mm}
\end{equation}
Here $b$ is the impact parameter for photons of finite rest mass
\cite{wheel}, and it is the distance perpendicular from the centre of the black hole to the
normal line on the ray of light intersecting the observer at
infinity \cite{Iyerp}.

The motion of geodesics is a force-free unaccelerated
motion. In the presence of a gravitational field, photons experience
gravitational force and this force comes due to the effective
potential. Here, the effective potential for photons travelling in
spacetime (\ref{1}) is given by
\begin{equation}\label{8}
  V_\text{eff}=\frac{L^{2}}{r^{2}}\Big(1-\frac{2M}{r}-\sigma r\Big).
\end{equation}
Note that the effective potential has different values of $\sigma$ for nonextreme, extreme and naked singularity of KBH, i.e., for
nonextreme  $0<\sigma<\frac{1}{8M}$,  for extreme $\sigma=\frac{1}{8M}$ while for naked singularity $\sigma>\frac{1}{8M}$. Here we discuss nonextreme and naked singularity cases and the extreme case will be discussed in Sec. V. When $\sigma=0$ then Eq. $(\ref{8})$ reduces
to Schwarzschild BH's effective potential, i.e.,
\begin{equation}\label{9}
  V^{S}_\text{eff}=\frac{L^{2}}{r^{2}}\Big(1-\frac{2M}{r}\Big).\\
\end{equation}

\begin{figure}[!ht]
\centering
\includegraphics[width=12cm]{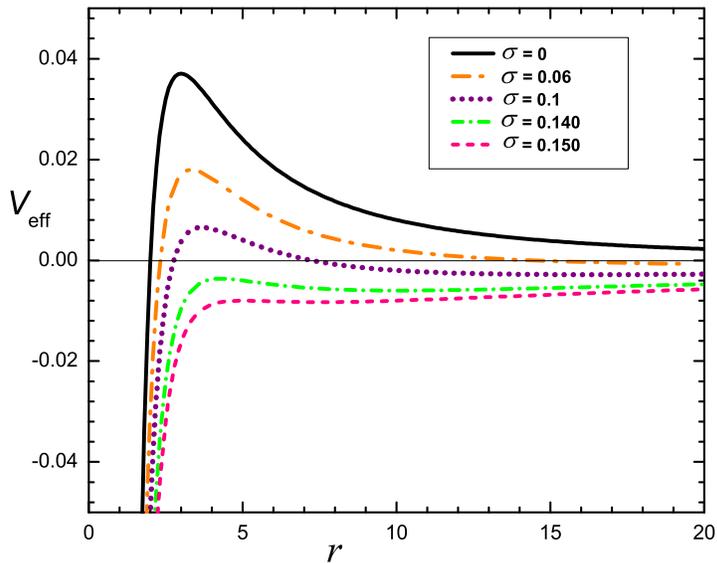}
\caption{Effective potential $V_\text{eff}$ of photons as a function
of distance $r$ from black hole, setting $M=1$. Top curve for
Schwarzschild black hole, middle two curves for nonextreme while
bottom two curves for naked singularity of KBH.} \label{ps}
\end{figure}

In Fig. $\ref{ps}$, the effective potential $V_\text{eff}$ is plotted
to study the behavior of photons near the considered
spacetime $(\ref{1})$ for different values of quintessence parameter
$\sigma$. We take $M=1$ for plotting $\sigma=\frac{1}{8}=0.125$ and the limits on $\sigma$ become for the nonextreme case $0<\sigma<0.125$, for the extreme case $\sigma=0.125$ (discussed later in Sec. V), and for naked singularity $\sigma>0.125$.
 Hence $\sigma=0$ corresponds to the Schwarzschild black hole, $\sigma=0.06$ and $0.1$
corresponds to the nonextreme KBH. For these cases photons do not cross
the horizon while at $\sigma=0.14$ and $\sigma=0.15$ photons cross
the horizon. In each curve there is no minima. Therefore, there is
no stable orbit for the photons, only an unstable orbit exists in
each case which corresponds to the maximum value $V_\text{max}$.

\section{Critical Variables And the Equation OF Path For Photons for KBH}

To find the radius of circular orbit of photons, we use the
condition  $\frac{dV_\text{eff}}{dr}=0$ to obtain
\begin{equation}\label{10}
  r_{c\pm}=\frac{1\pm\sqrt{1-6M\sigma}}{\sigma}.
\end{equation}
Here $r_{c+}$ is greater than the outer horizon $r_{+}$ while
$r_{c-}$ lies between the inner and outer horizons
$(r_{-}<r_{c-}<r_{+})$. The region of interest is between the
horizons. Therefore, the radius of an unstable circular orbit for
a photon is $r_{c-}=r_\text{ps}$, also called the photon sphere.
For the critical value of the photon sphere, conditions  imposed on
$\sigma$ are $0<\sigma<\frac{1}{8M}$ for the nonextreme and  $\sigma>\frac{1}{8M}$ for naked singularity. In the limit $\sigma\rightarrow
0$ we get the radius of photon sphere $r^{S}_\text{ps}=3M$ for the Schwarzschild black hole.
Now,  we convert the equation of motion
$(\ref{7})$ in terms of $u=\frac{1}{r}$.
We obtain the equation of path for photons
\begin{equation}\label{16}
\hspace{10mm} \Big(\frac{du}{d\phi}\Big)^{2}- B(u)= 0,
\end{equation}
where
\begin{equation}\label{17}
  B(u)=\frac{1}{b^{2}}-u^{2}\Big(1-2Mu-\frac{\sigma}{u}\Big).
\end{equation}
For critical value of the closest approach, we put $\frac{du}{d\phi}=0$
\cite{Ohani}. Identifying this point of the closest approach as $u=u_{2}$, from Eq. $(\ref{16})$, we have
\begin{equation}\label{18}
  \frac{1}{b^{2}}= u^{2}_{2}-2Mu_{2}^{3}-\sigma u_{2}.
\end{equation}
Substituting $u_{2}=\frac{1}{r_\text{ps}}$ from Eq.
$(\ref{10})$ in Eq. $(\ref{18})$, we obtain the critical value of
impact parameter for circular orbits
\begin{equation}\label{19}
  b_\text{sc}= \sqrt{\frac{r_\text{ps}^{3}}{r_\text{ps}-2M-\sigma r_\text{ps}^{2}}}.
\end{equation}
\newpage
The value of the impact parameter also imposes the same limits  on the
quintessence parameter $\sigma$, for both nonextreme and naked
singularity of KBH as mentioned above. For $\sigma=0$, Eq. $(\ref{19})$
gives the impact parameter $b^{S}_\text{sc}=3\sqrt{3}M$ for a Schwarzschild black hole.
 According to the circular orbit condition (setting $B(u)=0$) and solving  Eq. $(\ref{17})$, we
get one real root $u_{1}$ and two other roots $u_{2}$ and $u_{3}$,
$(u_{3}>u_{2}>u_{1})$ which are

\begin{equation}
\scriptsize{u_{1}= \frac{r_{o}-2M-\sqrt{(1-8M\sigma)r_{o}^{2}+4Mr_{o}-12M^{2}}}{4Mr_{o}},
~~~u_{2}=\frac{1}{r_{o}},
~~~u_{3}=\frac{r_{o}-2M+\sqrt{(1-8M\sigma)r_{o}^{2}+4Mr_{o}-12M^{2}}}{4Mr_{o}}}.
\end{equation}


Thus Eq. $(\ref{17})$ becomes
\begin{equation}\label{24}
  B(u)=2M(u-u_{1})(u-u_{2})(u-u_{3}).
\end{equation}
Substituting Eq. $(\ref{24})$ in $(\ref{16})$ yields
\begin{equation}\label{25}
  \frac{du}{d\phi}=\pm\sqrt{2M(u-u_{1})(u-u_{2})(u-u_{3})}.
\end{equation}
In Eq. (25), the positive sign $(+)$ shows that the angle $\phi$; changes more than $\pi$, that is the photon
	trajectory is bent toward KBH and for the negative sign $(-)$ the photon trajectory is bent away from KBH. For a ray of light, both $r_{o}$ and $b$ are obviously different from each other. Using Cardano's method
 solving the cubic equation,
\begin{equation}\label{26}
  r_{o}^{3}+\sigma b^{2}r_{o}^{2}-b^{2}r_{o}+2Mb^{2}=0,
\end{equation}
 the relation between $b$ and $r_{o}$ is
\begin{equation}\label{28}
  r_{o}=2\sqrt{\frac{\sigma^{2}b^{4}+3b^{2}}{9}}\cos\Big[\frac{1}{3}\cos^{-1}\Big(-\frac{2\sigma^{3}b^{6}+9\sigma b^{4}+54Mb^{2}}{6\sigma ^{2}b^{4}+18b^{2}}
  \sqrt{\frac{9}{\sigma^{2}b^{4}+3b^{2}}}\Big)\Big]-\frac{\sigma b^{2}}{3}.
\end{equation}
At $\sigma=0$, it consistently reduces to the Schwarzschild black hole lensing case
\cite{Iyerp},
\begin{equation}\label{29}
  r_{o}=\frac{2b}{\sqrt{3}}\cos\Big[\frac{1}{3}\cos^{-1}\Big(\frac{-3\sqrt{3}M}{b}\Big)\Big].
\end{equation}
\newpage

 \begin{figure}[!ht]
\centering
\includegraphics[width=12cm]{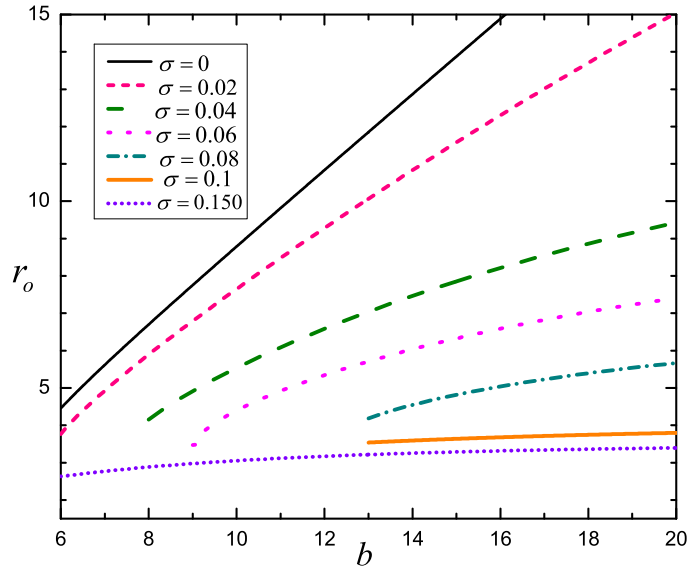}
\caption{Closest approach $r_{o}$ as a function of
impact parameter $b$ $(M=1)$. We discuss here the relation between
the closest approach $r_{o}$ and impact parameter $b$ for KBH lensing cases: -nonextreme and naked-singularity- and
compared it with a Schwarzschild black hole lensing case for different
values of $\sigma$.}\label{ps1}
\end{figure}

From Fig. $\ref{ps1}$, we observe that by increasing the value of
$b$, $r_{o}$ increases. In the region of the photon sphere $\sigma=[0,~0.1]$, $r_{o}$ depends on $b$ from the quintessence
parameter $\sigma$. Moreover, as $\sigma$
increases, light moves closer to KBH and the closest approach $r_{o}$
decreases. Therefore, $\sigma=0$ corresponds to a Schwarzschild black hole (taken as a
reference) while $\sigma=0.02$ to $\sigma=0.1$ correspond to the
nonextreme KBH. Beyond the photon sphere (region where no horizon
exists), i.e., $\sigma=0.150$, the light goes into the KBH, whereas $r_{o}$ remains constant and naked singularity occurs.

\section{Bending angle}

Suppose that a light ray comes from infinity $(\text{say}
-\infty)$, reaches the black hole at $r_{o}$, and finally moves back to infinity $(\text{say} +\infty)$ that is the observer. Due to
this  change, the angular coordinate $\phi$ is two times from
infinity to $r_{o}$. The light ray deflects from a straight line
path at the difference of $\pi$ which results in the bending angle
$\hat{\alpha}$  \cite{weinb}
\begin{equation}\label{30}
  \hat{\alpha}=2\int_{0}^{\frac{1}{r_{o}}}\frac{d\phi}{du}du-\pi.
\end{equation}
If we substitute Eq. $(\ref{25})$ into Eq. $(\ref{30})$, we obtain
\begin{equation}\label{31}
  \hat{\alpha}=2\int_{0}^{\frac{1}{r_{o}}}\frac{1}{\sqrt{2M(u-u_{1})(u-u_{2})(u-u_{3})}}du-\pi.
\end{equation}
If we write Eq. $(\ref{31})$ in terms of complete elliptic integral\footnote{The integral involving a rational function which contains square roots of cubic or quartic polynomials. Generally, here a definite cubic integrand that has a built-in command as \\$K(m)=F(\frac{\pi}{2}\mid m)=\int_{0}^{\frac{\pi}{2}}\frac{d\theta}{\sqrt{1-m\sin^{2}\theta}}$ } and an incomplete elliptic integral \footnote{If $\phi$ has the range $-\frac{\pi}{2}<\phi<\frac{\pi}{2}$ then $F(\phi\mid m) = \int_{0}^{\infty}\frac{d\theta }{\sqrt{1-m\sin^{2}\theta}}$.} we need to separate the integration limits into two parts:
\begin{equation}\label{32}
  \hat{\alpha}=\sqrt{\frac{2}{M}}\Big[\int_{u_{1}}^{\frac{1}{r_{o}}}\frac{1}{\sqrt{(u_{1}-u)(u-u_{2})(u_{3}-u)}}du
  -\int_{u_{1}}^{0}\frac{1}{\sqrt{(u_{1}-u)(u-u_{2})(u_{3}-u)}}du\Big]-\pi.
\end{equation}
Here the integrals can be recognized in terms of a first kind of
elliptical integral, where $u_{3}>u_{2}>u_{1}$ \cite{Byrdf}. Hence
\begin{equation}\label{33}
  \hat{\alpha}=2\sqrt{\frac{2}{M}}\Big[\frac{F(\Psi_{1},k)}{\sqrt{u_{3}-u_{1}}}-\frac{F(\Psi_{2},k)}{\sqrt{u_{3}-u_{1}}}\Big]-\pi.
\end{equation}
The integral variables can be defined as
 \begin{equation}\label{34}
   \Psi_{1}=\frac{\pi}{2}, ~~~~~ \Psi_{2}=\sin^{-1}\sqrt{\frac{r_{o}-2M-\sqrt{(1-8M\sigma)r_{o}^{2}+4Mr_{o}-12M^{2}}}{r_{o}-6M-\sqrt{(1-8M\sigma)r_{o}^{2}+4Mr_{o}-12M^{2}}}}.
 \end{equation}
In the elliptical integral modulus $k$ has a range $0\leq |k|^{2}\leq
1$, where
\begin{equation}\label{36}
  k = \sqrt{\frac{6M-r_{o}+\sqrt{(1-8M\sigma)r_{o}^{2}+4Mr_{o}-12M^{2}}}{2\sqrt{(1-8M\sigma)r_{o}^{2}+4Mr_{o}-12M^{2}}}}.
\end{equation}
Now $F(\frac{\pi}{2},k)\equiv K(k)$ defines a complete elliptical
integral while  $F(\Psi,k)$ is an incomplete elliptic integral. By
simplifying Eq. $(\ref{33})$, an exact bending angle can be obtained:
\begin{equation}\label{37}
  \hat{\alpha}= 4\sqrt{\frac{r_{o}}{\sqrt{(1-8M\sigma)r_{o}^{2}+4Mr_{o}-12M^{2}}}}\Big[K(k)-F(\Psi,k)\Big]-\pi.
\end{equation}
From the last expression, $\hat{\alpha}$ can be deduced for
nonextreme KBH under $0<\sigma<\frac{1}{8M}$ and for naked
singularity KBH under $\sigma>\frac{1}{8M}$. For $\sigma=0$, Eq. $(\ref{37})$, reduces to the Schwarzschild bending angle $\hat{\alpha}^{S}$ \cite{Iyerp}.
\newpage
 \begin{figure}[!ht]
\centering
\includegraphics[width=12cm]{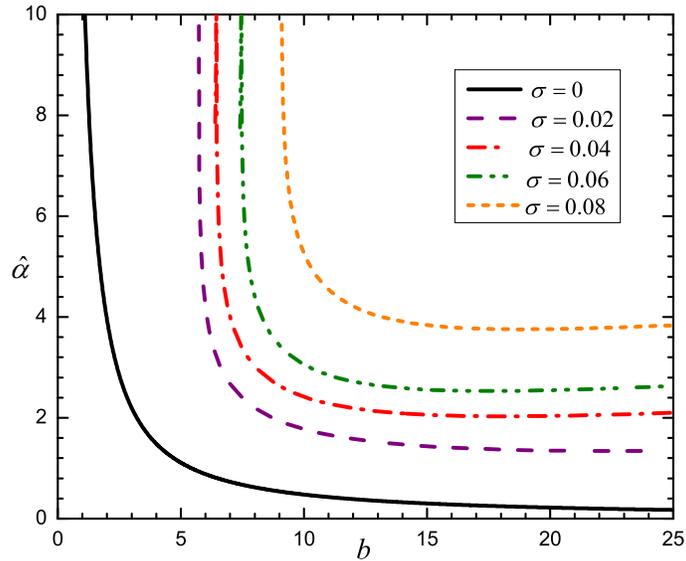}
\caption{Bending angle is a function of impact parameter $b$. This
is the case of nonextreme KBH lensing and its
maximum deflection value depends on the quintessence parameter
$0<\sigma<\frac{1}{8} ~(M=1)$. Here the Schwarzschild case occurs at $\sigma=0$
while $\sigma=0.02$ to $0.08$ for nonextreme case.} \label{ps2}
\end{figure}
Figure~\ref{ps2}, shows that the maximum deflection of light will
occur at the critical value of the impact parameter $b_\text{sc}$ in Eq.
$(\ref{17})$. Below $b_\text{sc}$ there will be no deflection and
above $b_\text{sc}$, we will get a continuous deflection (light circulates around the black
hole).
 Each single curve shows that by increasing the value of $b$, the bending angle decreases at different values of $\sigma$. Nevertheless, originally when we increases the value of $\sigma$, the critical value of the closest approach decreases since the light goes closer to the black hole. Similarly, the value of $b$ (near the photon sphere where maximum deflection occurs) decreases and the bending angle increases.\\
 
\begin{figure}[!ht]
\centering
\includegraphics[width=10cm]{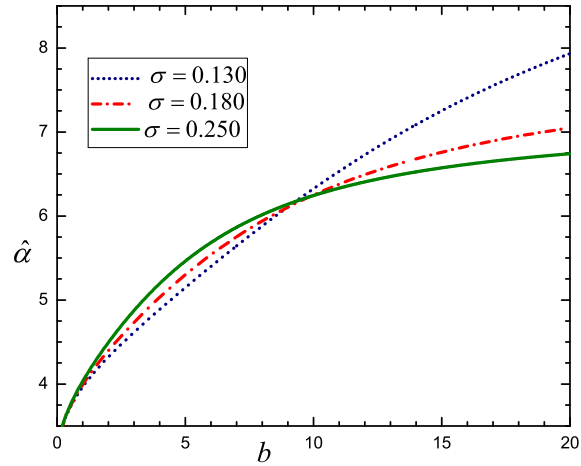}
\caption{Bending angle $\hat{\alpha}$ as a function of $b$ for naked singularity. At $M=1$, $\sigma>\frac{1}{8}$.} \label{ps3}
\end{figure}

\begin{figure}[!ht]
\centering
\includegraphics[width=10cm]{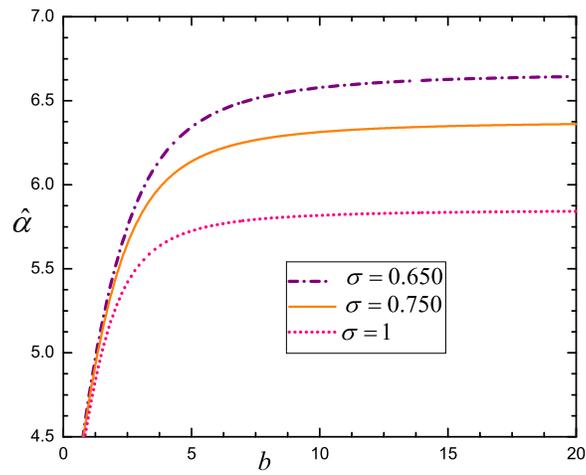}
\caption{$\hat{\alpha}$ as a function of $b$ for a naked singularity.} \label{ps3a}
\end{figure}

Figures~\ref{ps3} and \ref{ps3a} display the behavior of naked singularity. In Fig.~\ref{ps3}, for any curve at short distances, as $b$ increases the bending angle increases. In Fig.~\ref{ps3a}, for a long distance,  as $b$ increases the bending angle remains constant.
However, when we observe the whole phenomena, we see that the bending angle also depends on $\sigma$. As $\sigma$ increases, the bending angle decreases for both short and long ranges distances. Furthermore, when we compare the graph (Figs. $\ref{ps3}$ and $\ref{ps3a}$) of the naked singularity bending angle with the nonextreme and extreme bending angles graphs (Figs.~\ref{ps2} and \ref{ps6}), we observe that naked singularity behaves opposite from nonextreme and extreme cases.

\section{Gravitational Lensing by Extreme Kiselev Black Hole}
Extreme gravitational lensing is very amazing for some important phenomenona but it demands a great effort to be observed. In extreme gravitational lensing, where KBH is used as a lens, we need to discuss the bending of photons that pass very close to the lens and suffer a very large deflection.
\\
For the extreme Kiselev black hole (EKBH) we have $\sigma=1/8M$, thus the function $f(r)$ becomes
\begin{eqnarray}\label{38}
       f(r)=1-\frac{2M}{r}-\frac{r}{8M}.
\end{eqnarray}
This is an EKBH case for which $f(r)=0$ gives $r^{e}_\text{H}=4M$ known as a
degenerate solution (single horizon).
This value is twice the Schwarzschild black hole horizon, so it can be written as  $r^{e}_\text{H}=2r^{S}_\text{H}$.
Repeating the same procedure of Sec. II, for $\sigma=\frac{1}{8M}$ we obtain the
effective potential
\begin{equation}\label{44}
  V^{e}_\text{eff}=\frac{L^{2}}{r^{2}}-\frac{2ML^{2}}{r^{3}}-\frac{L^{2}}{8Mr},
\end{equation}
where the first term is related to the centrifugal potential. The second term represents the relativistic correction due to general relativity. The third term arises due to the fact that EKBH geometry depends on a parameter $\sigma=\frac{1}{8M}$. Due to the effect of this potential, we can see the behavior of a photon surrounding by the EKBH.
\begin{figure}[!ht]
\centering
\includegraphics[width=12cm]{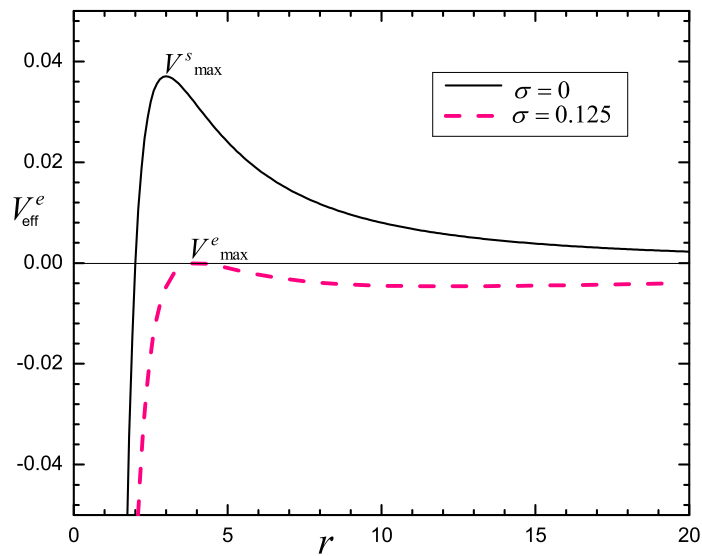}
\caption{Effective potential $V^{e}_\text{eff}$ is shown as a
function of distance $r$ taking for extreme Kiselev lensing
phenomenon. Observe that there is no minima (have no stable orbit) and only one maximum $V_\text{max}$, an unstable orbit that
exists which corresponds to $V^{e}_\text{max}$. Schwarzschild's
effective potential is taken as a reference
$(\sigma=0)$.}\label{ps4}
\end{figure}

\newpage
\section{Equation of Path And Critical values for EKBH}

Substituting $\sigma=\frac{1}{8M}$ in Eq. $(\ref{7})$, we obtain the first order nonlinear differential equation for path
\begin{equation}\label{48}
  \Big(\frac{du}{d\phi}\Big)^{2}- B^{e}(u)= 0,
\end{equation}
where
\begin{equation}\label{49}
  B^{e}(u)=\frac{1}{b^{2}}-u^{2}\Big(1-2Mu-\frac{1}{8M u}\Big).
\end{equation}

In Eq. $(\ref{49})$ we need o apply the circular orbit condition. This condition gives a cubic equation that has one real root $u_{1}^{e}<0$ and two
distinct positive roots such that $u^{e}_{3}>u^{e}_{2}>0$. The roots are

\begin{equation}\label{50}
  u^{e}_{1}= \frac{r^{e}_{o}-2M-2\sqrt{(r^{e}_{o}-3M)M}}{4Mr^{e}_{o}},
  ~~~~~~ u^{e}_{2}=\frac{1}{r^{e}_{o}}, ~~~~~~ u^{e}_{3}= \frac{r^{e}_{o}-2M+2\sqrt{(r^{e}_{o}-3M)M}}{4Mr^{e}_{o}}.
\end{equation}

Therefore, Eq. $(\ref{49})$ can be rewritten as
\begin{equation}\label{53}
  B^{e}(u)= 2M(u-u^{e}_{1})(u-u^{e}_{2})(u-u^{e}_{3}).
\end{equation}
If we replace again this equation into the equation of path, Eq. $(\ref{48})$, we obtain
\begin{equation}\label{54}
  \frac{du}{d\phi}= \pm \frac{1}{\sqrt{2M(u-u^{e}_{1})(u-u^{e}_{2})(u-u^{e}_{3})}}.
\end{equation}
In the limit $u=0$ $(r\rightarrow\infty)$, Eq. $(\ref{48})$ gives
\begin{equation}\label{55}
  u=\frac{\phi}{b}+constant.
\end{equation}
 For the critical value of the closest approach (radius of photon sphere  $r_{o}$), applying the second circular orbit condition $\frac{du}{d\phi}|_{u=\frac{1}{r_{o}}}=0$, and then the condition $\frac{dB^{e}(u)}{d\phi}|_{u=\frac{1}{r_{o}}}=0$ in Eq. $(\ref{48})$, we get
  $r^{e}_\text{c+}= 4M$ and $r^{e}_\text{c-}=12M$. Here, $r^{e}_\text{c+}=r^{e}_\text{H}$ gives a degenerate solution (with
$b=0$) whereas $r^{e}_\text{c-}=r^{e}_\text{ps}$ gives the  photon sphere. 
Now, by putting the value of $ b^{e}_\text{sc}$ into Eq. $(\ref{48})$ and using the
condition of circular orbit $B^{e}(u)=0$, we get the critical value
of the impact parameter, which is  $b^{e}_\text{sc}= 6\sqrt{6}M$.
For EKBH the relation between $r_{o}$ and $b$ is
\begin{equation}\label{61}
   r^{e}_{o}=\frac{b\sqrt{b^{2}+192M^{2}}}{12M}\cos\Big[\frac{1}{3}\cos^{-1}
   \Big\{-\frac{(b^{4}+288b^{2}+13824)}{b^{2}(b^{2}+192M^{2})^{\frac{3}{2}}}\Big\}\Big]
   -\frac{b^{2}}{24M}.
\end{equation}
\begin{figure}[!ht]
\centering
\includegraphics[width=11cm]{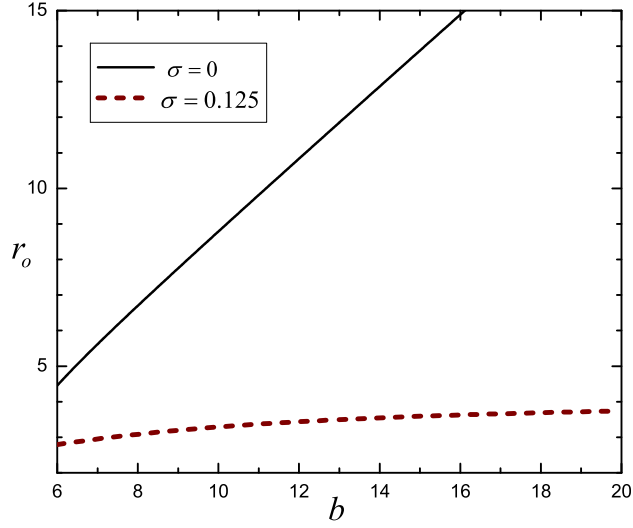}
\caption{Closest approach $r_{o}$ as a
function of the impact parameter $b$ for the EKBH. We see that by increasing the
value of the impact parameter $b$ the closest approach $r_{o}$
increases. Schwarzschild black hole case $(\sigma=0)$ is taken as reference
while for EKBH we take $\sigma=0.125$ with $M=1$.} \label{ps5}
\end{figure}
\newpage
\section{Bending Angle For Extreme Kiselev Black Hole }

The bending angle for the extreme Kiselev black hole (EKBH) can be obtained
by putting Eq. $(\ref{54})$ into $(\ref{30})$ where $r_{o}\rightarrow
r^{e}_{o}$. Doing this we obtain
\begin{equation}\label{62}
  \hat{\alpha}^{e}=2\int_{0}^{\frac{1}{r^{e}_{o}}}\frac{1}{\sqrt{2M(u-u^{e}_{1})(u-u^{e}_{2})(u-u^{e}_{3})}}du-\pi.
\end{equation}
We can decompose the limits and convert the integral into complete and
incomplete elliptical integral forms as follows
\begin{equation}\label{63}
  \hat{\alpha}^{e}=\sqrt{\frac{2}{M}}\Big[\int_{u^{e}_{1}}^{\frac{1}{r^{e}_{o}}}\frac{1}{\sqrt{(u^{e}_{1}-u)(u-u^{e}_{2})(u^{e}_{3}-u)}}du
  -\int_{u^{e}_{1}}^{0}\frac{1}{\sqrt{(u^{e}_{1}-u)(u-u^{e}_{2})(u^{e}_{3}-u)}}du\Big]-\pi.
\end{equation}
Both integrals can be recognized in terms of first kind of
elliptical integral \cite{Byrdf}, where the integrand has the condition
$u^{e}_{3}>u^{e}_{2}>u^{e}_{3}$. Thus we have
\begin{equation}\label{64}
  \hat{\alpha}^{e}=\sqrt{\frac{2}{M}}\Big[\frac{2F(\Psi^{e}_{1},k^{e})}{\sqrt{u^{e}_{3}-u^{e}_{1}}}-\frac{2F(\Psi^{e}_{2},k^{e})}
  {\sqrt{u^{e}_{3}-u^{e}_{1}}}\Big]-\pi.
\end{equation}
Simplification of Eq. $(\ref{64})$ gives
\begin{equation}\label{65}
  \hat{\alpha}^{e}=4\sqrt{\frac{2r^{e}_{o}}{\sqrt{(r^{e}_{o}-3M)M}}}\Big[\frac{F(\Psi^{e}_{1},k^{e})}{\sqrt{u^{e}_{3}-u^{e}_{1}}}-\frac{F(\Psi^{e}_{2},k^{e})}
  {\sqrt{u^{e}_{3}-u^{e}_{1}}}\Big]-\pi.
\end{equation}
For EKBH, elliptic integral parameters can be defined
as
\begin{equation}\label{66}
   \Psi^{e}_{1}=\frac{\pi}{2}, ~~~~~~ \Psi^{e}_{2}=\sin^{-1}\sqrt{\frac{r^{e}_{o}-2M-2\sqrt{(r^{e}_{o}-3M)M}}{r^{e}_{o}-6M-2\sqrt{(r^{e}_{o}-3M)M}}}.
\end{equation}
Modulus $k^{e}$ has range $0\leq |k^{e}|^{2}\leq 1$, where
\begin{equation}\label{68}
  k^{e} = \sqrt{\frac{6M-r^{e}_{o}+2\sqrt{(r^{e}_{o}-3M)M}}{4\sqrt{(r^{e}_{o}-3M)M}}}.
\end{equation}
Thus, the exact bending angle for EKBH lensing is given by
\begin{equation}\label{69}
  \hat{\alpha}^{e}= 2\sqrt{\frac{2r_{o}}{\sqrt{(r^{e}_{o}-3M)M}}}\Big[K(k^{e})-F(\Psi^{e},k^{e})\Big]-\pi,
\end{equation}
where $F(\frac{\pi}{2},k^{e})\equiv K(k^{e})$ defines the complete
elliptical integral and $F(\Psi^{e},k^{e})$ is an incomplete
elliptical integral.
\begin{figure}[!ht]
\centering
\includegraphics[width=12cm]{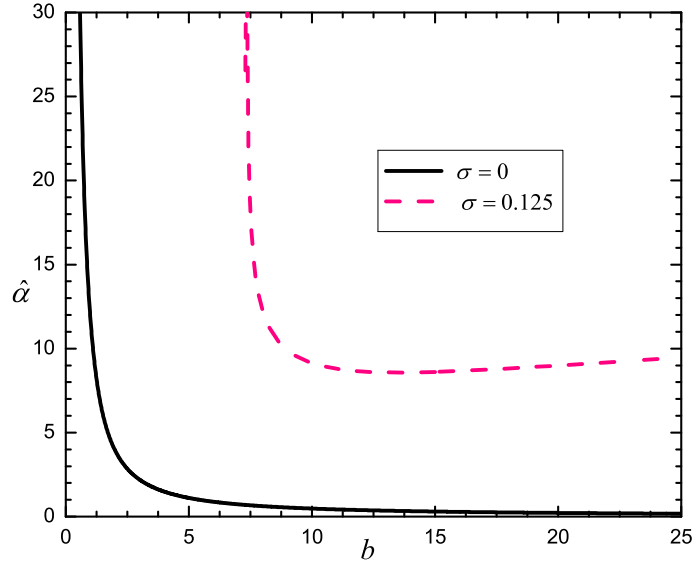}
\caption{For extreme Kiselev black hole lensing, the bending angle
$\hat{\alpha}^{e}$ is a function of the impact parameter $b$ (setting $M=1$). In
this case, the bending angle also depends on the value of the quintessence
parameter $\sigma$. In this figure, $\sigma=0.125$ is the value for the extreme case while $\sigma=0$ is for the Schwarzschild black hole bending angle taken as a reference.} \label{ps6}
\end{figure}

Figure~\ref{ps6}, shows that by increasing the value of $b$, the bending
angle decreases. The dashed curve shows the bending
angle for EKBH, while the solid curve shows the bending angle for the Schwarzschild black hole. Both curves display the same behavior since they have one horizon. In EKBH lensing, the event horizon is twice the
Schwarzschild's horizon $(r^{S}_{H})$. However, the difference
between these two bending angles is that in the extreme case, the bending angle is larger than the Schwarzschild black hole bending angle because if we increase the value of the quintessence parameter $\sigma$, the bending angle will also increase.
\section{Alternative Approach For Finding Bending Angle}

Gravitational lensing phenomena involves the study of the null geodesic equations. When the solution of the space-time geometry
$(\ref{1})$ extends, an event horizons exist at $r_{+}$ and $r_{-}$, see Eq. $\ref{2}$. Our main interest is in the region that lies between the horizons, which  is called the photon sphere $r_{ps}$ [Eq. $\ref{10}$]. Therefore, the deflection will occur when a ray of light passes through that region with the closest approach $r_{o}$. In order to compute the bending angle $\hat{\alpha}$ we need to compute the value of the impact parameter $b$. If we divide Eq. $(\ref{6})$ with $(\ref{7})$ we obtain
\begin{equation}\label{70}
  \frac{d\phi}{dr}=\frac{1}{r^{2}\sqrt{\frac{1}{b^{2}}-\frac{1}{r^{2}}\Big(1-\frac{2M}{r}-\sigma r\Big)}}.
\end{equation}
Now, for the closest approach $r=r_{o}$ and
$\frac{dr}{d\phi}|_{r=r_{o}}=0$, we will have
\begin{equation}\label{71}
  b(r_{o})=\frac{r_{o}}{\sqrt{1-\frac{2M}{r_{o}}-\sigma r_{o}}}.
\end{equation}
By substituting Eq. $(\ref{71})$ in Eq. $(\ref{70})$, we obtain
\begin{equation}\label{72}
  \frac{d\phi}{dr}=\frac{1}{r\sqrt{\Big(\frac{r}{r_{o}}\Big)^{2}\Big(1-\frac{2M}{r_{o}}-\sigma r_{o}\Big)-\Big(1-\frac{2M}{r}-\sigma r\Big)}}.
\end{equation}
We adopt the procedure of \cite{weinb}, thus we will use the following bending angle
formula:
\begin{equation}\label{73}
  \alpha=2\int_{r_{o}}^{\infty} \frac{d\phi}{dr}dr-\pi.
\end{equation}
By using Eq. $(\ref{72})$, the deflection angle for a light ray
becomes
\begin{equation}\label{74}
  \alpha(r_{o})=2\bigints_{r_{o}}^{\infty}{\frac{dr}{r\sqrt{\Big(\frac{r}{r_{o}}\Big)^{2}\Big(1-\frac{2M}{r_{o}}-\sigma r_{o}\Big)-
  \Big(1-\frac{2M}{r}-\sigma r\Big)}}}-\pi.
\end{equation}
\begin{figure}[!ht]
\centering
\includegraphics[width=10cm]{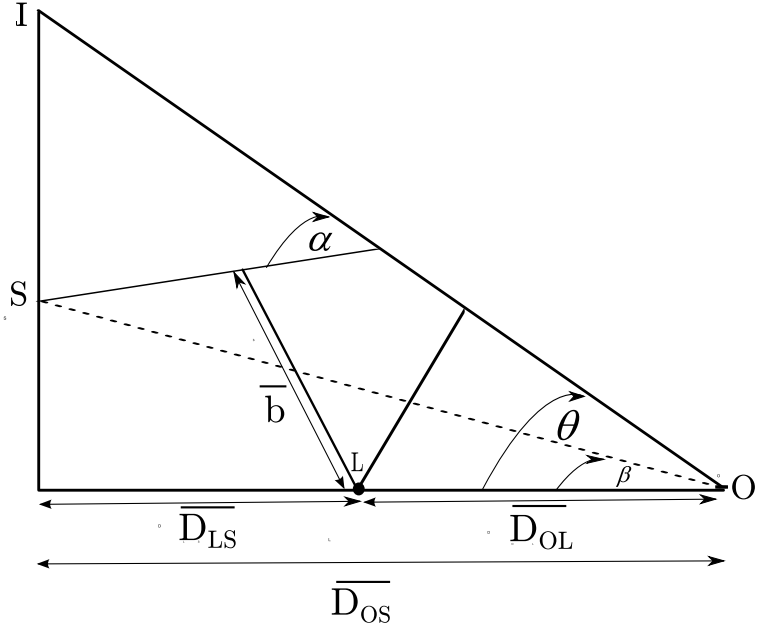}
\caption{The lens diagram. The positions of observer (O), source
(S), lens (L) and image (I) are shown.
  The observer-lens, observer-source, and lens-source distances are represented by $\overline{D_\text{OL}}, \overline{D_\text{OS}}$ and $ \overline{D_\text{LS}}$, respectively.} \label{LD}
\end{figure}\\
The geometry of a lensing phenomenon is shown in Fig.
$\ref{LD}$. This figure is commonly called ``the lens diagram".  The lens equation can be expressed as \cite{Virbh}
\newpage
\begin{equation}\label{75}
  \tan\beta=\tan\theta-\frac{\overline{D_\text{LS}}}{\overline{D_\text{OS}}}\Big[\tan(\alpha-\theta)+\tan\theta\Big],
\end{equation}
where $\overline{D_\text{LS}}$ is the distance from the lens to the source and $\overline{D_\text{OS}}$  is the distance from the observer to the source. We also have
\begin{equation}\label{76}
  b(r_{o})=\overline{D_\text{OL}}\sin\theta,
\end{equation}
where $\overline{D_\text{OL}}$ is the distance from the observer to the lens. Angular positions of source and images are represented by $\beta$
and $\theta$, respectively while the deflection angle due to a black
hole
is denoted by $\alpha$ as it is shown in Figure~\ref{LD}.\\
Now, if we convert the distance and the impact parameter in terms of the
Schwarzschild black hole radius, we find
\begin{eqnarray}\label{77}
   X&=& \frac{r}{2M},~~~~~~X_{o}=\frac{r_{o}}{2M},~~~~~~b(r_{o})=2Mb(X_{0}),
\nonumber\\
\nonumber\\
  d_\text{ol}&=&\frac{\overline{D_\text{OL}}}{2M},~~~~~d_\text{os}=\frac{\overline{D_\text{OS}}}{2M},~~~~~~~d_\text{ls}=\frac{\overline{D_\text{LS}}}{2M}.
\end{eqnarray}
From here, we will introduce a new quintessence parameter $\sigma_{\ell}=2M\sigma$ in terms of the Schwarzschild radius.
Using Eqs. $(\ref{76})$ and $(\ref{77})$ in Eqs. $(\ref{74})$,
$(\ref{71})$, $(\ref{2})$, and $(\ref{10})$ respectively, we get
\begin{eqnarray}
  \alpha(X_{o})&=&2\bigints_{X_{o}}^{\infty}\frac{dX}{X\sqrt{\Big(\frac{X}{X_{o}}\Big)^{2}\Big(1-\frac{1}{X_{o}}-\sigma_{\ell}X_{o}\Big)
  -\Big(1-\frac{1}{X}-\sigma_{\ell}X\Big)}}-\pi,\label{79}\\
  b(X_{o})&=&\frac{X_{o}}{\sqrt{1-\frac{1}{X_{o}}-\sigma_{\ell}X_{o}}}=d_\text{ol}\sin\theta,\label{80}\\
X_\text{H}&=&\frac{1}{2\sigma_{\ell}} \pm \frac{1}{\sigma_{\ell}}\sqrt{\frac{1}{4}-\sigma_{\ell}}, ~~~~~~~ X_\text{ps}=\frac{1-\sqrt{1-3\sigma_{\ell}}}{\sigma_{\ell}}, \label{81}
\end{eqnarray}
where $X_\text{H}$ denotes the distance from the horizons and $X_\text{ps}$ is the distance from the photon sphere.\\
In order to find the position of images, we need to solve Eq. $(\ref{75})$ for the source position
$\beta$ along with Eqs. $(\ref{79})$ and $(\ref{80})$.
\par
Generally, for a circular symmetric lens, the magnification is given
by \cite{Virbh}
\begin{equation}\label{83}
  \mu=\Big|\frac{\sin\beta}{\sin\theta}\frac{d\beta}{d\theta}\Big|^{-1}.
\end{equation}
 Here, the tangential magnifications and the radial magnifications are respectively defined as
\begin{equation}\label{84}
  \mu_\text{t}\equiv \Big(\frac{\sin\beta}{\sin\theta}\Big)^{-1},~~~~~~~~\mu_\text{r}\equiv \Big(\frac{d\beta}{d\theta}\Big)^{-1}.
\end{equation}
By differentiating both sides of Eq. $(\ref{75})$, we get
\cite{Ernes}
\begin{equation}\label{85}
  \frac{d\beta}{d\theta}=\Big(\frac{\cos\beta}{\cos\theta}\Big)^{2}\Big[1-\frac{d_\text{ls}}{d_\text{os}}\Big\{1+
  \Big(\frac{\cos\theta}{\cos(\alpha-\theta)}\Big)^{2}\Big(\frac{d\alpha}{d\theta}-1\Big)\Big\}\Big],
\end{equation}
where
$\frac{d\alpha}{d\theta}=\frac{d\alpha}{dX_{o}}\frac{dX_{o}}{d\theta}$.
By taking the derivative of Eq. $(\ref{79})$ with respect to $X_{o}$, we
obtain
\begin{equation}\label{86}
  \frac{d\alpha}{dX_{o}}=\bigints_{X_{o}}^{\infty}\frac{X\Big(2X_{o}-3-\sigma_{\ell}X_{o}^{2}\Big)   }{2X_{o}^{4}\Big[\Big(\frac{X}{X_{o}}\Big)^{2}\Big(1-\frac{1}{X_{o}}-\sigma_{\ell}X_{o}\Big)
      -\Big(1-\frac{1}{X}-\sigma_{\ell}X\Big)\Big]^{\frac{3}{2}}}dX.
\end{equation}
Finally, by differentiating Eq. $(\ref{72})$ with respect to $\theta$ on both
sides and doing some simplifications we get
\begin{equation}\label{87}
  \frac{dX_{o}}{d\theta}=\frac{X_{o}\Big(1-\frac{1}{X_{o}}-\sigma_{\ell}X_{o}\Big)^{\frac{3}{2}}\sqrt{1-(\frac{X_{o}}{d_\text{ol}})^{2}(1-\frac{1}{X_{o}}-\sigma_{\ell}X_{o})^{-1}}}
  {\frac{1}{2d_\text{ol}}\Big(2X_{o}-3-\sigma_{\ell}X_{o}^{2}\Big)}.
\end{equation}

\section{Weak Field Limit}

We are going to take some approximations in this section. If the source and the
lens are aligned,  then we can approximate $\tan\beta\approx \beta$ and
$\tan\theta\approx \theta$. For the relativistic images we can write
$\Delta\alpha=2n\pi+\Delta\alpha_{n}$, (where $n$ is an integer) and
$0<\Delta\alpha_{n}\leq1$. Hence, we can replace
$\tan(\alpha-\theta)$ by $\Delta\alpha_{n}-\theta$. If the ray of
light reaches the observer after it turns around the black hole, the
deflection angle $\alpha$ must be very close to $2\pi$. Therefore,
Eq. (\ref{75}) becomes
\begin{equation}\label{88}
  \beta=\theta-\frac{\overline{D_\text{LS}}}{\overline{D_\text{OS}}}~\Delta\alpha_{n}=\theta-\frac{d_\text{ls}}{d_\text{os}}~\Delta\alpha_{n},
\end{equation}
and the impact parameter is $b=d_\text{ol}\theta$.

Relativistic images are formed only if the ray of light passes very
close to the photon sphere. For the closest approach $X_{o}$, it is
convenient to write
\begin{equation}\label{90}
X_{o}=X_\text{ps}+\varepsilon ~~~~~~~~\Big(0\leq \varepsilon\ll1\Big).
\end{equation}
For a Schwarzschild black hole, the approximated deflection angle will be \cite{Bozza}
\begin{equation}\label{91}
  \alpha\sim-2\ln\Big[\frac{2+\sqrt{3}}{18}\varepsilon\Big]-\pi.
\end{equation}
Therefore, we shall also look for a similar approximation
\cite{Ernes}
\begin{equation}\label{92}
  \alpha=-A\ln(B \varepsilon)-\pi,
\end{equation}
where $A$ and $B$ are positive numbers that we take from
\cite{Ernes}. However, in our case these numbers will depend only on $\sigma_{\ell}$. Therefore, we will have
\begin{equation}\label{93}
  A= \lim_{X_{o}\rightarrow X_\text{ps}}\Big[-\Big(X_{o}-X_\text{ps}\Big)\frac{d\alpha_\text{exact}}{dX_{o}}\Big],
  ~~~~   B=\lim_{X_{o}\rightarrow X_\text{ps}}\Big[\frac{\exp\Big\{\Big(-\frac{\alpha_\text{exact}+\pi}{A}\Big)\Big\}}{\Big(X_{o}-X_\text{ps}\Big)}\Big].
\end{equation}
Now, by taking the value of $X_{o}$ from Eq. $(\ref{90})$ and by putting that expression into Eq.
$(\ref{80})$, we get the impact parameter in terms of $\varepsilon$ as
\begin{equation}\label{95}
  b(X_\text{ps}+\varepsilon)=\frac{X_\text{ps}+\varepsilon}{\sqrt{1-\frac{1}{X_\text{ps}+\varepsilon}-\sigma_{\ell}\Big(X_\text{ps}+\varepsilon\Big)}}.
\end{equation}
If we use a Taylor expansion in $\varepsilon$ up to second order in Eq. $(\ref{95})$, we get the impact parameter as
\begin{equation}\label{96}
  b=C-D\varepsilon^{2}, \hspace{6mm}
\end{equation}
where
\begin{eqnarray}
C&=&\frac{\Big(1-\sqrt{1-3\sigma_{\ell}}\Big)^{\frac{3}{2}}}{\sigma_{\ell}\sqrt{-1+\sqrt{1-3\sigma_{\ell}}+2\sigma_{\ell}}},\label{97}\\
 D&=&\frac{\sigma_{\ell}\Big\{-\Big(2-2\sqrt{1-3\sigma_{\ell}}\Big)+\sigma_{\ell}\Big(8-5\sqrt{1-3\sigma_{\ell}}-6\sigma_{\ell}\Big)\Big\}}
 {2\Big(-1+\sqrt{1-3\sigma_{\ell}}+2\sigma_{\ell}\Big)^{2}\sqrt{-2+2\sqrt{1-3\sigma_{\ell}}+5\sigma_{\ell}-2\sigma_{\ell}\sqrt{1-3\sigma_{\ell}}}}.\label{98}
\end{eqnarray}
Now we can find the value of $\varepsilon$ from Eq. $(\ref{96})$ using
\begin{equation}\label{99}
  \varepsilon=\sqrt{\frac{C-b}{D}}. \hspace{20mm}
\end{equation}
Finally, If we substitute the value of $\varepsilon$ [Eq. $(\ref{99})$ into $(\ref{92})$], we obtain the approximated bending angle
expression
\begin{equation}\label{100}
  \alpha=-A \ln\Big[B\sqrt{\frac{C-d_\text{ol}\theta}{D}}\Big]-\pi.
\end{equation}

\section{Relativistic Images}

Virbhadra and Ellis defined ``relativistic images'' of a
gravitational lens as those images which occur due to light
deflections by angles $\hat\alpha>3\pi/2$ \cite{Virbh}. Similarly,
when $\beta=0$ and $\hat\alpha>2\pi$, the location of relativistic
``Einstein rings" are specified \cite{viru}. For a fixed value of
$\beta$, we can get $\theta$ related to the positions of
corresponding images. Thus, we can do an approximation using a first order
Taylor expansion around $\alpha=2n\pi$ for the position of the nth
relativistic image \cite{Ernes}
\begin{equation}\label{101}
  \theta \approx \theta_{n}^{o}-\rho_{n}\Delta \alpha_{n},
\end{equation}
where $\theta=\theta_{n}^{o}$ at $\alpha=2n\pi$ and
\begin{equation}\label{102}
  \rho_{n}=-\frac{d\theta}{d\alpha}\mid_{\alpha=2n\pi}.
\end{equation}
For the value of $\theta$ we take Eq. $(\ref{100})$, and we get
\begin{eqnarray}
  \theta&=&\frac{1}{d_\text{ol}}\Big[C-\frac{D}{B^{2}}\exp\Big\{\frac{-2}{A}\Big(\alpha+\pi\Big)\Big\}\Big],\label{103}\\
   \theta_{n}^{o}&=&\frac{1}{d_\text{ol}}\Big[C-\frac{D}{B^{2}}\exp\Big\{\frac{-2}{A}\Big(2n+1\Big)\pi\Big\}\Big].\label{104}
\end{eqnarray}
Taking derivatives in ($\ref{104}$) and then substituting into ($\ref{102}$), we obtain
\begin{equation}\label{105}
  \rho_{n}= -\frac{1}{d_\text{ol}}\Big[\frac{2D}{AB^{2}}\exp\Big\{\frac{-2}{A}\Big(2n+1\Big)\pi\Big\}\Big].
\end{equation}
From Eq. $(\ref{101})$, we have
\begin{equation}\label{106}
  \Delta\alpha_{n} \approx \frac{\theta_{n}-\theta_{n}^{o}}{-\rho_{n}}.
\end{equation}
Using Eqs. $(\ref{104})$ and $(\ref{105})$ in $(\ref{106})$, we
get
\begin{equation}\label{107}
  \Delta\alpha_{n} \approx \frac{A}{2}\Big[\Big\{\frac{d_\text{ol}B^{2}}{D}
  \exp\Big\{\frac{2}{A}\Big(2n+1\Big)\pi\Big\}\Big\}\theta_{n}-\Big\{\frac{B^{2}C}{D}\exp\Big\{\frac{2}{A}\Big(2n+1\Big)\pi\Big\}-1\Big\}\Big].
\end{equation}
Substituting Eq. $(\ref{106})$ into $(\ref{88})$ yields
\begin{equation}\label{108}
  \beta=\theta_{n}-\frac{d_\text{ls}}{d_\text{os}}\Delta\alpha_{n}.
\end{equation}
Putting Eq. $(\ref{107})$ into $(\ref{108})$, we get
\begin{equation}\label{109}
  \beta=\Big[1+\frac{d_\text{ls}d_\text{ol}}{d_\text{os}}\Big\{\frac{AB^{2}}{2D}\exp\Big\{\frac{2}{A}\Big(2n+1\Big)\pi\Big\}\Big\}\Big]\theta_{n}
-\frac{d_\text{ls}}{d_\text{os}}\Big[\frac{A}{2}\Big\{\frac{B^{2}C}{D}\exp\Big\{\frac{2}{A}\Big(2n+1\Big)\pi\Big\}-1\Big\}\Big].
\end{equation}
In order to obtain the approximate position for the relativistic images, we
neglect the number $1$ because $(\frac{d_\text{ls}d_\text{ol}}{d_\text{os}} \gg 1)$ in this approximation. Therefore, we have
\begin{equation}\label{110}
 \theta_{n}=\frac{d_\text{os}}{d_\text{ls}d_\text{ol}}\Big[\frac{2D}{AB^{2}}\exp\Big\{\frac{-2}{A}\Big(2n+1\Big)\pi\Big\}\Big]\beta+\frac{1}{d_\text{ol}}\Big[C-\frac{D}{B^{2}}\exp\Big\{\frac{-2}{A}\Big(2n+1\Big)\pi\Big\}\Big].
\end{equation}
Here in Eq. $(\ref{110})$,  if the source, lens, and image are perfectly
aligned then $\beta=0$ and we can obtain the Einstein ring with angular radius
\begin{equation}\label{111}
  \theta_{n}^{E}=\frac{1}{d_\text{ol}}\Big[C-\frac{D}{B^{2}}\exp\Big\{\frac{-2}{A}\Big(2n+1\Big)\pi\Big\}\Big]=\theta_{n}^{o}.
\end{equation}
The amplification of the nth relativistic image is given by
\begin{equation}\label{112}
  \mu_{n}\approx \Big|\frac{\beta}{\theta_{n}} \frac{d\beta}{d\theta_{n}}\Big|^{-1}.
\end{equation}
Tangential magnification for relativistic images is
\begin{equation}\label{113}
  \mu_{t}= \frac{\theta_{n}}{\beta}=\frac{d_\text{os}}{d_\text{ls}d_\text{ol}}\Big[\frac{2D}{AB^{2}}\exp\Big\{\frac{-2}{A}\Big(2n+1\Big)\pi\Big\}\Big]
  +\frac{1}{\beta d_\text{ol}}\Big[C-\frac{D}{B^{2}}\exp\Big\{\frac{-2}{A}\Big(2n+1\Big)\pi\Big\}\Big].
\end{equation}
Radial magnification for relativistic images is
\begin{equation}\label{114}
  \mu_{r}= \frac{d\theta_{n}}{d\beta}=\frac{d_\text{os}}{d_\text{ls}d_\text{ol}}\Big[\frac{2D}{AB^{2}}\exp\Big\{\frac{-2}{A}\Big(2n+1\Big)\pi\Big\}\Big].
\end{equation}
Thus, the total amplification of the nth relativistic images can be
calculated by combining both tangential magnification Eq.
$(\ref{113})$ and radial magnification Eq. $(\ref{114})$ in
$(\ref{112}$), which yields
\begin{equation}\label{115}
  \mu_{n}= \frac{1}{|\beta|}\frac{d_\text{os}}{d_\text{ls}d_\text{ol}}\Big[\frac{2D}{AB^{2}}\exp\Big\{\frac{-2}{A}\Big(2n+1\Big)\pi\Big\}\Big]
  \Big[\frac{1}{d_\text{ol}}\Big\{C-\frac{D}{B^{2}}\exp\Big\{\frac{-2}{A}\Big(2n+1\Big)\pi\Big\}\Big\}\Big].
\end{equation}
Here, if the observer, lens and source are aligned $(\beta=0)$,  the
amplification will diverge. Therefore, the size of the relativistic
images will become very small and the brightness will be low. For the total
magnification of relativistic images, the sum of the relativistic image is taken into account
\begin{equation}\label{116}
  \mu_{R}=2\Sigma_{n=1}^{\infty} \mu_{n}=\frac{2}{|\beta|}\frac{d_\text{os}}{d_\text{ls}}\Sigma_{n=1}^{\infty} \theta_{n}^{o}\rho_{n}.
\end{equation}
Now, by using the geometric series $\Sigma_{n=1}^{\infty}
a^{n}=\frac{a}{1-a}$ for $|a| <1$, the total
magnification of the relativistic images will be
\begin{equation}\label{117}
  \mu_{R} \approx \frac{2}{|\beta|} \frac{d_\text{os}}{d_\text{ls}d_\text{os}^{2}} \frac{2D}{AB^{2}}
  \Big[\frac{D}{B^{2}}
   \Big\{\frac{\exp\Big({-12\pi/A}\Big)}{1-\exp\Big({-8\pi/A}\Big)}\Big\}-C\Big\{ \frac{\exp\Big({-6\pi/A}\Big)}{1-\exp\Big({-4\pi/A}\Big)}\Big\}\Big].
\end{equation}

\section{Discussion}
We have studied the GL scenario for nonextreme, naked singularity and extreme cases for KBH. We discussed the null geodesics for these three cases in order to study the behavior of the scalar field. We observed that effective potential and the null-geodesics trajectories depend on the quintessence parameter. From Figs.~\ref{ps} and~\ref{ps4}, we found that the potential does not have a minimum value so there is no stable circular orbit for photons. Moreover there are only unstable orbits for all cases. We also studied the behavior of the light in the lensing process of KBH. From Figures \ref{ps1} and \ref{ps5}, we ensured that as the value of impact parameter $b$ is increased the value of $r_{0}$ increases. We have worked with the quintessence field, so due to the effect of quintessence parameter $\sigma$, the situation gets reversed i.e., closest approach $r_{0}$
decreases by increasing the value of $b$ and light goes closer to the KBH. Moreover, when $\sigma$ reaches to $0.125$, the $r_{0}$ remains constant with respect to $b$. For this, we calculated the equation of the path and the bending angle $\hat{\alpha}$. After that, we converted this expression in terms of elliptic integrals. The bending angle depends on the value of $\sigma$. For each case, $\sigma$ has different limits. We solved the elliptical integrals numerically and  studied their behavior via plots in Figs. $\ref{ps2}$, $\ref{ps3}$, $\ref{ps3a}$, and  $\ref{ps6}$.

We also studied a GL phenomenon for nonextreme KBH $(0<\sigma<\frac{1}{8M})$. In this case, it can be seen from Fig. $\ref{ps2}$, that as the value of the impact parameter increases, the bending angle decreases. Nevertheless, for the whole process, for large value of $\sigma$, light goes closer to the black hole and the bending angle would be larger. Furthermore, when we compared it with the Schwarzschild case, we observed that $\hat{\alpha^{S}}$ is smaller than the bending angle for the nonextreme case.

For a GL phenomenon for EKBH, we have $\sigma=\frac{1}{8M}$. From Fig.~\ref{ps6}, we noticed that as the impact parameter $b$ increases, the bending angle $\hat{\alpha}^{e}$ for EKBH  decreases. When we compared it with Schwarzschild black hole, we observed that its behavior is similar to the Schwarzschild black hole bending angle $\hat{\alpha}^{S}$ and nonextreme bending angles, since EKBH has only one horizon which is twice the Schwarzschild's horizon.  However, $\hat{\alpha}^{e}$ is greater than the $\hat{\alpha}^{S}$.

To study GL phenomena for naked singularity, we took $\sigma>\frac{1}{8M}$.  In this case, the behavior of the light is totally different as there is no horizon and the value of the closest approach $r_{o}$ will remain constant with respect to $b$. From Fig. $\ref{ps3}$, it can be seen that as we increase the value of $b$, the bending angle increases. However, from Figs. \ref{ps3} and \ref{ps3a}, one can conclude that the bending angle is smaller for large $\sigma$. For the case of naked singularity, we found that the bending angle is larger than the nonextreme, extreme and Schwarzschild cases. (The order of the bending angles is naked singularity $>$ extreme KBH $>$ nonextreme KBH $>$ Schwarzschild black hole) Additionally, the behavior of a naked singularity bending angle is almost opposite both nonextreme KBH and extreme KBH bending angles. We calculated the bending angle by another approach in Sec. VIII and we found that the results are similar for both approaches. We have also calculated the approximated bending angle by using the weak field limit. The expression for the magnification of relativistic images are also derived.

One can generalize this analysis and comparison for the Reissner-Nordstr\"{o}m black hole surrounded by quintessence matter and the study of relativistic images can also be done more rigorously. This type of work might be important for studying the highly redshifted galaxies, quasars, supermassive black holes, exoplanets, dark matter candidates and so on.
	
\section{Acknowledgment}
	
	The authors would like to thank K. S. Virbhadra for insightful
	comments on this work. S.B. is supported by the Comisi\'on Nacional de Investigaci\'on
Cient\'ifica y Tecnol\'ogica (Becas Chile Grant No. 72150066). M.J. and A.Y. are supported via NRPU Grant No. 20-2166 from Higher Education Commission Islamabad.

\end{document}